

\font\twelverm=cmr10 scaled 1200    \font\twelvei=cmmi10 scaled 1200
\font\twelvesy=cmsy10 scaled 1200   \font\twelveex=cmex10 scaled 1200
\font\twelvebf=cmbx10 scaled 1200   \font\twelvesl=cmsl10 scaled 1200

\font\twelvett=cmtt10 scaled 1200   \font\twelveit=cmti10 scaled 1200
\font\twelvesc=cmcsc10 scaled 1200  
\skewchar\twelvei='177   \skewchar\twelvesy='60
\def\twelvepoint{\normalbaselineskip=12.4pt plus 0.1pt minus 0.1pt
  \abovedisplayskip 12.4pt plus 3pt minus 9pt
  \belowdisplayskip 12.4pt plus 3pt minus 9pt
  \abovedisplayshortskip 0pt plus 3pt
  \belowdisplayshortskip 7.2pt plus 3pt minus 4pt
  \smallskipamount=3.6pt plus1.2pt minus1.2pt
  \medskipamount=7.2pt plus2.4pt minus2.4pt
  \bigskipamount=14.4pt plus4.8pt minus4.8pt
  \def\rm{\fam0\twelverm}          \def\it{\fam\itfam\twelveit}%
  \def\sl{\fam\slfam\twelvesl}     \def\bf{\fam\bffam\twelvebf}%
  \def\mit{\fam 1}                 \def\cal{\fam 2}%
  \def\sc{\twelvesc}               \def\tt{\twelvett}
  \def\sf{\twelvesf}
  \textfont0=\twelverm   \scriptfont0=\tenrm   \scriptscriptfont0=\sevenrm
  \textfont1=\twelvei    \scriptfont1=\teni    \scriptscriptfont1=\seveni
  \textfont2=\twelvesy   \scriptfont2=\tensy   \scriptscriptfont2=\sevensy
  \textfont3=\twelveex   \scriptfont3=\twelveex  \scriptscriptfont3=\twelveex
   \textfont\itfam=\twelveit
  \textfont\slfam=\twelvesl
  \textfont\bffam=\twelvebf \scriptfont\bffam=\tenbf
  \scriptscriptfont\bffam=\sevenbf
  \normalbaselines\rm}
\def\journal#1 #2 #3 #4.{{\it #1} {\bf #2}, #3 (#4).}
 \twelvepoint
\baselineskip=20pt
\parindent 20pt
\settabs 4 \columns
\parskip=10pt
\hfuzz=2pt   
\def\line{\hbox to \hsize}
\def\frac #1#2{{#1\over #2}}

\def\psid{\psi^{\dagger}}
\def\Phid{\Phi^{\dagger}}

\def\ad{ a^{\dagger}}

\def\sgn{{\rm sgn\,}}

\def \r{{\bf r}}

\def \z{{\bar z}}

\def \ket #1{{\vert #1\rangle}}
\def \bra #1{{\langle #1\vert}}


\line{\hfil NSF-ITP-94-15}
\line{\hfil February 1994}
\vskip 1cm
\centerline{\bf LAUGHLIN STATES AT}
\centerline{\bf  THE EDGE }
\vskip 1cm

\centerline{Michael Stone\footnote{$^\dagger$}{
                        \it Permanent Address:
                         University of Illinois at Urbana Champaign,
                         Loomis Laboratory of Physics,
                         1110 W. Green St.,
                         Urbana, IL 61801.}}
\centerline{\&}
\centerline{Matthew P.~A.~Fisher}
\vskip .5cm

\centerline{\it Institute for Theoretical Physics}
\centerline{\it University of California, Santa Barbara}
\centerline{\it Santa Barbara, CA 93106}
\centerline{\it USA}
\line{\hfil}

\line{\bf Abstract\hfil}

An effective wavefunction for the edge excitations in the
Fractional quantum Hall effect can be found by dimensionally
reducing the bulk wavefunction. Treated this way the Laughlin
$\nu=1/(2n+1)$ wavefunction yields  a Luttinger model ground
state. We  identify the edge-electron field with a Luttinger
hyper-fermion operator, and the edge electron itself with a
non-backscattering Bogoliubov quasi-particle. The edge-electron
propagator may be  calculated directly from the effective
wavefunction using the properties of a one-dimensional
one-component plasma, provided a prescription is adopted which is
sensitive to the extra flux attached to the electrons.

\vfil

\eject

\line{\bf 1. Introduction\hfil}

In introducing his chiral Luttinger liquid theory for the edge
states in the fractional quantum Hall effect (FQHE), Wen [1] gave
a persuasive, but indirect, argument for Luttinger-like behaviour
of the  the edge-electron Green functions. For  droplets composed
of electrons in the simplest FQHE  phases with filling fraction
$\nu=1/m$ ($m$ an odd integer)  he concluded that there should be
only one branch of edge excitations, and that the edge-electron
propagator should decay with a power law
$$
\bra{droplet} T\{\psid(x,t)\psi (0,0)\}\ket{droplet} \propto
\frac 1{(x-vt)^{\xi}},
\eqno (1.1)
$$
where $v$ is the velocity of the
unidirectional edge waves, and $x$ the distance along the
circumference. The exponent $\xi$ turns out to be
equal to $m$.  The identification of the FQHE edge with a
Luttinger liquid  opens up a number of possibilities for
confronting  theory  with experiment. For example, the problem of
resonant tunneling between edge states can be mapped [2] onto
previous work on one-dimensional electron gasses [3] and this
gives results that  agree  quite well with experiment [4].

Despite this success there are still some aspects of the
FQHE/Luttinger liquid correspondence that could be more
transparent. As in the conventional Luttinger liquid [5] the
exponent $\xi$ is determined by a singularity in the
single-particle occupation-number density at the Fermi surface.
For a circular droplet the relation $\xi=m$  implies that the
occupation of lowest Landau level orbitals with angular momentum
$N$ must go to zero as $(N_{max}-N)^{m-1}$. That this is true for
the Laughlin wavefunction has been confirmed  by numerical
evaluation of the occupation numbers for small numbers of
electrons [6], and by an analytic calculation of the density
using the one-component plasma interpretation [7]. To the best of
our knowledge  however the Luttinger liquid behaviour of the edge
states has  not been directly connected to properties of the FQHE
wavefunction. This paper is  intended to provide such a link.

A relation between the edge correlator and the bulk FQHE
wavefunction can be found by exploring the analogy between the
conventional, non-chiral, Luttinger liquid and the $\nu=1/m$ FQHE
phases. This analogy goes further than the sharing of power law
correlators. Both systems have Jastrow type wavefunctions
composed of products of differences of particle coordinates.
The  product wavefunction proposed by Laughlin for the FQHE  [8]
has a large overlap with the true ground state, and more
importantly,  precisely reproduces its long-distance part
[9,10,11]. The ground-state wavefunction for the Luttinger (or
Thirring)  model is also known to be of Jastrow type [12]. It
coincides with that of the Sutherland model [13] whose
correlators were already known to  be described by the same
conformal field theory as the Luttinger-Thirring model [14].

Given the similarity of the correlators, can it be merely a
coincidence that the  wavefunctions of these systems also
resemble one another?  It is true that one is a product of
coordinate differences in {\it one\/} dimension while the other
is a product of differences in {\it two\/} dimensions, and that
one system has only unidirectional (chiral) excitations while the
other has modes propagating in both directions.  Nonetheless,
despite the difference of dimension, we expect the  similarity of
the wavefunctions to be more than an accident. The large magnetic
field effectively halves the number of dimensions   and  converts
the lowest Landau level into a phase-space with $x=$ `$q$', $y=$
`$p$'. This  dimensional reduction leads to  a very close analogy
between the edge of a QHE droplet and  a Fermi surface. The
incompressible bulk of the QHE fluid corresponds to the Fermi
sea, and heading inward from the edge to diving to deeper
momentum.

For a {\it full\/} Landau level at least, we can make this
dimensional reduction precise.  By freezing  the radial
coordinates at the droplet radius, and treating the resulting
wavefunction  as  one for   free, one-dimensional electrons,  we
can recover most of the edge physics [15]. Unfortunately  this
simple recipe for creating an  edge effective wavefunction does
not extend straightforwardly to the fractional states. If we
adjust the coupling constant in the Luttinger-Thirring model so
as to give  a Laughlin-like one-dimensional  wavefunction with
factors $(z_i-z_j)^m$, the Luttinger model fermion propagator
does not have the desired $\xi=m$ exponent.  Perhaps this is due
to the very different dynamics of the Luttinger model?  For
non-vanishing coupling the left and right going Luttinger
particles mix with each other, so there may be a profound
difference between the chiral and non chiral systems. It turns
out that this fear is groundless and the recipe can be made to
work.

What has gone wrong is that  we have misidentified the
one-dimensional operator corresponding to the edge electron. In
the Luttinger-Thirring model there is a family of  operators
which create charge $e$ particles with fermi-like statistics, but which in
some sense acquire a phase $e^{i\pi m}$, ($m$ odd) when
interchanged [16]. The first member of this family  is the
fundamental Fermi field in the relativistic Thirring model. In
non-relativistic systems the operators  with $m>1$ are
responsible for the appearance of subdominant terms in the
two-point functions  arising from the discreteness of the
underlying charges.  We will see that the FQHE edge-electron operator
must be identified with one of these $m\pi$ statistics fields.
This hyper-fermion is not simply related to the fundamental
Luttinger fermions but can be written as an exponential of one of
the non-backscattering Bogoliubov quasi-particles. Since the edge
electrons do not backscatter in the absence of inter-edge
tunneling, this shows that the identification can be used for
dynamical properties, and not just for ground-state expectation
values.

In section two we will give more details of the edge correlators
and their connection to one-dimensional systems. In section three
we will review some of the properties of the Luttinger-Thirring
model and identify   the edge fermion with an appropriate Haldane
hyper-fermion operator. In section four we will use a
one-component plasma  method due to  Hellberg and Mele
[17] to calculate the  edge correlator directly from
the Laughlin wavefunction. This calculation will show how the two
dimensionality of the Hall system requires us to distinguish
between $e^{i\pi}$ and $e^{3i\pi}$. For completeness, the
appendix contains a derivation of the ground state wavefunction
for the Luttinger-Thirring model.

\vskip 40pt
\line{\bf 2. Edge correlators \hfil}

There is an obvious similarity between the expression for the
wavefunction of a droplet of $N$ lowest Landau level electrons at
filling fraction $\nu=1$
$$
\Psi_{\nu=1}(z_1,z_2,\ldots,z_N)=\prod_{i<j} (z_i-z_j)e^{-\frac 14 \sum
z_i^2}
\eqno (2.1)
$$
and the wavefunction of a set of $N$ one-dimensional,
non-relativistic, fermions filling the $N$ lowest energy plane wave states on a
ring of circumference $L$
$$
\Psi (x_1,x_2,\ldots, x_N)=e^{-ik_f\sum x_i}\prod_{i<j} (e^{2\pi
ix_i/L}-e^{2\pi ix_j/L}).
\eqno( 2.2)
$$
($k_f=\pi (N-1)/L$ if $N$  is odd).  In particular, apart from
trivial normalization and Fermi-momentum factors, the wavefunctions
coincide at the boundary of the droplet where $|z|=R=\sqrt{2N}$,
$z=Re^{i\theta}$,  and   $x=R\theta$ is the distance along
the circumference.

This similarity extends to the correlators. The field
operator for electrons in the lowest Landau level  is
$$
\psi(z)=
\sum_{n=0}^{\infty}
\hat a_n\frac 1{\sqrt{2\pi 2^n n!}}z^ne^{-\frac 14\vert
z\vert^2},
\eqno (2.3)
$$
where the $\hat a_n$ are fermionic annihilation operators obeying
$\{\hat \ad_n,\hat a_{n'}\}=\delta_{nn'} $. We can evaluate its
equal-time
two-point function --- {\it i.e} the one-particle density matrix
$$
\bra{\nu=1} \psid(z)\psi (z')\ket{\nu=1} =G(z,z').
\eqno (2.4)
$$
Away from the boundary of the droplet $G(z,z')$ decays as a gaussian
with a range of the magnetic length
$$
|G(z,z')|=\frac 1{2\pi}e^{-\frac 14 |z-z'|^2},
\eqno (2.5)
$$
but when both $z,z'$ approach the boundary, $G(z,z')$ becomes
long ranged. Explicitly
$$
G(x,x')=e^{-iN(x-x')/R} \int_{-\infty}^\infty
{d\xi}\sqrt{\frac
B{2\pi}}e^{-B\xi^2/2}\frac 1{(x-x')-\xi+i\epsilon}.
\eqno (2.6)
$$
Once again  $x$ denotes  distance along the boundary (assumed
small compared to the circumference), and we have temporarily
restored the magnetic field $B$ to make it manifest that (2.6) is
a one-dimensional free-fermion correlator convoluted with a
factor which serves merely to  smooth it on scales shorter than a
magnetic length. Actually we find only  part of the usual $\sin
k_fx$ prefactor, but if we were to consider an annulus rather
than a disc we would find a  contribution  from the other fermi
point on letting $z,z'$ approach the inner edge.

Can this dimensional reduction continue to work when we
replace the $\nu=1$ droplet with a Laughlin $\nu=1/m$ state
$$
\Psi_{\nu=1/m}(z_1,z_2,\ldots,z_N)=
\prod_{i<j} (z_i-z_j)^me^{-\frac 14 \sum
z_i^2}\quad ?
\eqno (2.7)
$$
A one-dimensional wavefunction that resembles (2.7) is
the Luttinger-Thirring model ground state. This wavefunction is
(see ref. [12] and the appendix)
$$
\Psi_{\{\lambda\}}(x_1,x_2,\ldots, x_N)=\prod_{i<j} \sin
\frac \pi L(x_i-x_j)\left |\sin \frac \pi L
(x_i-x_j)\right|^{\lambda-1},
\eqno(2.8)
$$
where $\lambda$ depends on the interaction.
Perhaps this  wavefunction is not immediately obviously a
dimensionally reduced Laughlin state.  For $\lambda=m$ an odd
integer,  however, a seemingly  innocent manoeuver allows us to  write
$\Psi_{\{\lambda=m\}}$ as
$$
\Psi_{\{\lambda=m\}} (x_1,x_2,\ldots, x_N)=
{\rm const.}e^{-i(N-1)\pi/L\sum x_i}
\prod_{i<j} (e^{2\pi i
x_i/L}-e^{2\pi ix_j/L})^{m}.
\eqno(2.9)
$$
Once we drop the trivial  $e^{-i(N-1)\pi/L\sum x_i}$ factor  this
is clearly of Laughlin form.  We will use the notation
$$
\Psi_{m}= \prod_{i<j} (e^{2\pi i
x_i/L}-e^{2\pi ix_j/L})^{m}
\eqno (2.10)
$$
for this state.

The dynamics of the Luttinger model is quite different from the
FQHE edge. The edge has only a single branch of  excitations, but
the Luttinger fermion operator  couples to  both left and right
goers, and so we  must be prepared for  both $x+v_ft$ and
$x-v_ft$ dependence in the Green functions. Because of this we
will   focus initially on  equal-time correlators since these
depend only on ground state properties, and not on the spectrum.

Compare the integral
$$
G_2(z,z')=Z_2^{-1}\int d^2z_1\ldots d^2z_N
\prod_i(\z-\z_i)^{m}\prod_i(z'-z_i)^{m}
\prod_{i<j}|z_i-z_j|^{2m}e^{-\frac 12\sum_i |z|^2},
\eqno(2.11)
$$
which gives the FQHE one-particle density matrix, with that
giving the same quantity in the Luttinger model
$$
G_1(x,x')= Z_1^{-1}\int dx_1\ldots dx_N
\Psi^*_{m}(x,x_1,\dots, x_N)
\Psi_{m}(x',x_1,\ldots,x_N).
\eqno(2.12)
$$
Motivated by 2.6 we might conjecture that they coincide on the
boundary of the droplet
$$
G_2(Re^{i x/R},Re^{ix'/R})\buildrel ?\over =G_1(x,x').
\eqno (2.13)
$$
What evidence is there for this? Not much!  The
Fermion equal-time correlator for the Luttinger wavefunction
(2.8) is [14]
$$
\bra {\Psi_{\{\lambda\}}} \psid(x)\psi(x')\ket{\Psi_{\{\lambda\}}}
=\frac 1{(x-x')^{\frac 12
(\lambda+\frac 1\lambda)}}
\eqno (2.14)
$$
which does not reduce to (1.1) once $\lambda=m>1$. The conjecture
as stated cannot be true.

The heart of the problem  lies in the seemingly innocent rewriting of
(2.8) as (2.9).  The two wavefunctions are algebraically
identical, but in some sense the first  changes by a factor of
$e^{i\pi}$ under interchange of $x_i$ with $x_j$,  while the
second changes by a factor of $e^{im \pi}$ (m, as always, is
odd). Of course if  we were {\it strictly} in one dimension this
difference must be invisible. The FQHE particles however move in
the plane and do not pass directly through one another, so
they {\it can} perceive  the distinction.

How can we build this phase into the evaluation of the Luttinger
correlator? In the next section we will show that it may taken
into account  by altering the statistics of the operator we use
to describe the fermion.

\vskip 40pt
\line{\bf 3.  Haldane Fermions\hfil}

The Luttinger model on an interval of period $2\pi$ is defined by
the hamiltonian
$$
{\cal H}=\int dx \left\{ \frac 12 J_R^2+\frac 12 J_L^2+ \frac
g\pi J_RJ_L\right\}.
\eqno(3.1)
$$
Here $J_R, J_L$ are the currents for the left and right going
fermions. They obey
$$
[J_R(x), J_R(x')]=-[J_L(x), J_L(x')]=
-\frac{i}{2\pi}\partial_x\delta(x-x').
\eqno (3.2)
$$

The interaction may be  decoupled by introducing
a new set of currents $\tilde J_L, \tilde J_R$  defined by
$$
\eqalign{
J_R=&\cosh \alpha \tilde J_R+\sinh \alpha \tilde J_L \cr
J_L=&\sinh \alpha \tilde J_R+\cosh \alpha \tilde J_L .\cr
}
\eqno(3.3)
$$
If we set $-\tanh\alpha=g/\pi$ and express ${\cal H}$ in terms of
$J_{L,R}$ the cross term
disappears and
$$
{\cal H}={{\rm sech}\,} 2\alpha
\int dx \left\{ \frac 12 \tilde J_R^2+\frac 12 \tilde J_L^2\right\}.
\eqno (3.4)
$$

Both sets of currents obey the same commutation relations and
are formal conjugates  of each other --- {\it i.e\/} we can write
down a formal expression for a   unitary operator $U$ such that
$J_{R,L}=U^\dagger\tilde J_{R,L}U$. As usual, $U$ is a proper
unitary transformation only in a theory with a cutoff.

We can write the currents as derivatives of two independent
chiral boson fields
$$
J_R=\frac 1{2\pi} \partial_x \varphi_R
\qquad
J_L=\frac 1{2\pi} \partial_x \varphi_L,
\eqno (3.5)
$$
with
$$
[\varphi_R(x), \varphi_R(x')]=-[\varphi_L(x), \varphi_L(x')]=
{i\pi}\sgn(x-x'),
\eqno (3.6)
$$
and then bosonized expressions for  fundamental fermions in the system are
$$
\psi_R=:e^{i\varphi_R}: \qquad \psi_L=:e^{-i\varphi_L}:.
\eqno (3.7)
$$
We calculate their  correlators by introducing new $\tilde
\varphi_{L,R}$ in the same manner as $\tilde J_{R,L}$
$$
\eqalign{
\varphi_R=&\cosh \alpha \tilde \varphi_R+\sinh \alpha \tilde \varphi_L \cr
\varphi_L=&\sinh \alpha \tilde \varphi_R+\cosh \alpha \tilde \varphi_L .\cr
}
\eqno(3.8)
$$
The $\tilde \varphi_{R,L}$ are independent {\it free\/} fields
so substituting (3.8) in (3.7) allows us to compute correlators.
For example,
$$
\langle \psid_R(x)\psi_R(x')\rangle=\frac 1{(x-x')^{\cosh
2\alpha}}.
\eqno (3.9)
$$
This coincides with (2.12) after one identifies  $\lambda$ with
$e^{-2\alpha}$. This identification is confirmed by the
computation of other Luttinger-Thirring correlators.

To define the higher statistics hyper-fermion operators    we
follow Haldane [16] and define two new fields
$$
\eqalign{
\theta (x)=&\frac 12 (\varphi_R(x)+\varphi_L(x))\cr
\varphi(x)=&\frac 12 (\varphi_R(x)-\varphi_L(x)),
}
\eqno(3.10)
$$
which obey
$$
[\varphi(x),\varphi(x')]=[\theta(x),\theta(x')] =0
$$
$$
[\varphi(x),\theta(x')]=\frac i2\pi \sgn(x-x').
\eqno (3.11)
$$
We then define
$$
\Phi_m(x)=:e^{i\varphi(x)+im\theta(x)}:
\eqno (3.12)
$$
For arbitrary $m$ the operators in (3.12)
change the total charge $J_R+J_L$ by one unit.

Clearly $\Phi_1(x)=\psi_R(x)$ and $\Phi_{-1}(x)=\psi_L(x)$. From
(3.11) we see that
$$
\Phi_m(x)\Phi_m(x')=e^{im\pi \sgn(x-x')}\Phi_m(x')\Phi_m(x),
\eqno (3.13)
$$
so the $\Phi_m$ have Fermi statistics if $m$ is an odd integer and Bose
statistics if $m$ is even.

The $\Phi_m$ are what we need to make the connection between the
Luttinger model and the FQHE. It is they, not the fundamental
fermions, whose equal time correlator  corresponds to
(2.11).  We easily compute
$$
\langle \Phid_m(x)\Phi_m(x')\rangle =\frac 1{(x-x')^{\frac 12
(\lambda+\frac {m^2}{\lambda})}}.
\eqno (3.14)
$$
{\it Now\/} when we set $\lambda=m$ we find that
$$
\langle \Phid_m(x)\Phi_m(x')\rangle_{\lambda=m} =\frac
1{(x-x')^{m}}.
\eqno (3.15)
$$
This  coincides with (1.1) and supports  our identification of
the FQHE edge electron with the Luttinger hyper-fermion operator.

For $\lambda=e^{-2\alpha}=m$ we also find that
$$
\Phi_m(x)=:e^{i\sqrt {m}\tilde \varphi_R(x)}:
\eqno( 3.16)
$$
showing that the hyper-fermion operator is the exponential of a
field that, like the FQHE edge electron,  does not suffer
left/right mixing. This has the important {\it dynamical\/}
consequence that $\Phi_m$ depends only on $x-v_ft$, and  so the
FQHE edge-electron {\it propagator \/} and not just the equal-time
function  coincides with the Luttinger correlator.
We can use the other operator $\Phi_{-m}=:\exp -i\sqrt {m}\tilde
\varphi_L:$ to represent the electrons on the other side of a
Hall bar, or inner edge of an annulus.

Let us compare  with Wen's construction  of the FQHE edge-electron
field [1]. He uses the known velocity of the excitations to argue
that the commutator of the operators $J_{edge}$ measuring the
edge-electron number
must obey (see also [18])
$$
[J_{edge}(x), J_{edge}(x')]=\frac {-i}{2\pi m}\partial_x\delta(x-x').
\eqno( 3.17)
$$
He then defines an edge-boson field $\varphi_{edge}$ via
$$
\frac 1{2\pi m}\partial_x \varphi_{edge}(x)=J_{edge}(x),
\eqno (3.18)
$$
and identifies the edge electron  with
$$
\psi_{edge}(x)=:e^{i\varphi_{edge}(x)}:,
\eqno (3.19)
$$
since it changes the edge charge by unity, and has (hyper)-fermi
statistics. Clearly then we should identify $\varphi_{edge}$ with
$\sqrt{m}\tilde\varphi_{R}(x)$, and the edge charge, $J_{edge}$,
with $\tilde J_R(x)/\sqrt{m}$. In terms of the original
Luttinger-Thirring fields
$$
J_{edge}(x)=\frac 12 (J_R+J_L)+\frac 1{2m}(J_R-J_L).
\eqno (3.20)
$$
The edge-electron operator  $\Phi_m$ changes $\rho=(J_R+J_L)$ by
unity, and $j=(J_R-J_L)$ by $m$, the combination resulting in a
change of unity in $J_{edge}$.

Let us consider the significance of these operator
correspondences for using the FQHE system as a paradigm for
Luttinger liquid phenomenology. If we consider a real
one-dimensional system, electrons in a mesoscopic wire for
example, it may, under the right circumstances be modeled as a
Luttinger liquid [3]. Consider then a  point scattering impurity with a
potential $V(x)=V_0\delta(x)$. This  contributes a term
$$
{\cal H}_{imp}=
V_0\left(\psid_L(0)\psi_R(0)+\psid_R(0)\psi_L(0)\right)
\eqno (3.21)
$$
to the effective Luttinger model. Here  ${\cal H}_{imp}$ scatters
the fundamental Luttinger fermions from one Fermi point to the
other. This scattering is in addition to left-right mixing
produced by the $J_LJ_R$ term in (3.1).  If we want to deconvolve these
two processes we should express everything in terms of the
$\tilde \varphi_{L,R}$.  The bosonized version of (3.21) is
$$
{\cal H}_{imp}=V_0 :\cos 2\theta(0):.
\eqno (3.22)
$$
This is most naturally  expressed in terms of some new operators
$ \tilde \psi_{L,R}$ defined by
$$
\eqalign{
\tilde \psi_R=\left(\Phi_m\right)^{\frac 1m}\quad &\buildrel {\rm def} \over
= \quad :e^{i\varphi/m+i\theta}: \quad =\quad :e^{i\tilde
\varphi_R/\sqrt{m}}:\cr
\tilde \psi_L=\left(\Phi_{-m}\right)^{\frac 1m} \quad &\buildrel {\rm def}\over
= \quad :e^{i\varphi/m-i\theta}:\quad=\quad :e^{-i\tilde\varphi_L/\sqrt{m}}:,
\cr }
\eqno (3.23)
$$
as
$$
{\cal H}_{imp} =V_0\left (\tilde \psid_L(0)\tilde \psi_R(0) + {\rm
h.c}\right).
\eqno (3.24)
$$
In the absence of the impurity these two operators do not mix
left and right moving excitations.

The operators $\tilde \psid_{L,R}$ create an excitation with
charge $e/m$ and statistics $\pi /m$. When $m$ is an odd integer
these can be identifies with  the Laughlin quasi-particle on the
two edges of the FQHE fluid (in a Luttinger liquid $m$ need
{\it not\/} be an integer.). The operator ${\cal H}_{imp}$ then
hops a Laughlin quasi-particle from one edge of the FQHE to the
other. Even in the ordinary Luttinger  liquid a perturbation
expansion in $V_0$ describes backscattering of quasi-particles of
charge $e/m$ at each order in $V_0$.

\vskip 40pt
\line {\bf 4. A one-dimensional one-component Plasma \hfil }

In this section we return to the problem of finding the
edge correlator directly from the Laughlin wavefunction. We know
that naive dimensional reduction of the bulk wavefunction does
not give the correct exponent.  In the last section we saw that a
solution was to keep the one-dimensional approximation, but  to
modify the Luttinger operators. Now we must answer the question
of  how to obtain  the correlators  of these modified operators
from  the one-dimensional Luttinger wavefunction. We should also
ask  whether the procedure makes physical sense. We find that it
does when we  use a prescription based on a  method due to
Hellberg and Mele [17].

The Luttinger model one-particle density matrix is given by  the integral
$$
G_1(x,x')=\int dx_1\ldots dx_N
\prod_i (\z-\z_i)^m (z'-z_i)^m \prod_{i<j}\left\vert z_i-z_j\right\vert^{2m}.
\eqno (4.1)
$$
where $z=e^{2\pi i x/L}$, and similarly $z'$.
We begin by considering a simpler problem
$$
e^{-F(x,x')}= |z-z'|^{\frac m2} \int dx_1\ldots dx_N
\prod_i  |z-z_i|^m |z'-z_i|^m \prod_{i<j} |z_i-z_j|^{2m}.
\eqno (4.2)
$$
Finding $F$ is equivalent to determining  the force between  a pair of
logarithmically interacting charges of magnitude $1/\sqrt{2}$ inserted into a
gas of  similar charges of magnitude $\sqrt {2}$ which are
confined to a circular loop of length $L$.  The inverse
temperature of the gas is $m$.  The  factor outside the integral
is the mutual potential of the two charges.

If  we assume that the one-component plasma completely screens
the test charges at large distance, then  $F$ will become
independent of the distance and
$$
\int dx_1\ldots dx_N
\prod_i  |z-z_i|^m |z'-z_i|^m \prod_{i<j} |z_i-z_j|^{2m}
\propto \frac 1{|z-z'|^{m/2}}.
 \eqno (4.3)
$$
This is the correct exponent for correlator of  Luttinger
bosons (see ref. [14] eq. 18).

The integral  (4.1) giving the density matrix   differs
from (4.2) in containing extra phase factors. For the FQHE system
we want to pick up a factor of $e^{\pm im\pi}$ for each particle
lying between $x$ and $x'$. In a genuinely  one-dimensional
problem the choice of sign would pose a problem. In two
dimensions the ambiguity is resolved by the geometry. When  we
require $z,z'$ to lie on the outer edge of a Hall droplet the
electrons at $z_i$ will always be passed on their right as $z'$ circles the
droplet
counterclockwise, so the phase should increase by $m\pi$ each time
$z'$ passes by a $z_i$. If we were at the inner edge of an
annulus then we would select the opposite sign.

Using this insight, the FQHE density matrix can be written
$$
G^{\{m\}}_1(x,x')= e^{im \pi  n_0 (x'-x)}\frac 1{|x-x'|^{m/2}}
\left \langle e^{im\pi \int_x^{x'} \delta
n(\xi)d\xi}\right \rangle,
\eqno (4.4)
$$
where the angular brackets denote an expectation value for the
same Coulomb gas and $\delta n(\xi)$ is the excess number-density
over its mean value $n_0$. The superscript $m$ indicates that
we are using the $+m\pi$ phase recipe.
By introducing a new field $\chi$ with
$\delta n(x)=\partial_x\chi$ we can write this as
$$
\left \langle e^{im\pi \int_x^{x'} \delta
n(\xi)d{\xi}}\right \rangle =\int d[\chi(x)] e^{m\int
\partial_x\chi(x)\ln |x-x'|\partial_{x'}\chi(x') dxdx'+
im\pi(\chi(x')-\chi(x))}
\eqno (4.5)
$$
The gaussian functional integral is easily performed by going to
fourier space and using
$$
\int_{-\infty}^{\infty} \frac {dk}{|k|}e^{ik(x-x')}=
-2\ln |x-x'| + {\rm Constant.}
\eqno (4.6)
$$
We find
$$
G^{\{m\}}_1(x,x')\propto  e^{im \pi  n_0 (x'-x)}\frac 1{|x-x'|^{m/2}}
\frac 1{|x-x'|^{m^2/2m}}= e^{im \pi  n_0 (x'-x)}\frac 1{|x-x'|^{m}}.
\eqno (4.7)
$$
We have at last reproduced the desired exponent for the edge electron.

If we had taken the  phase factor for the passage of $z'$ past one
of the other particles as being $e^{i\pi}$ only, we would have
found instead
$$
G_1^{\{1\}}\propto  e^{i \pi  n_0 (x'-x)} \frac
1{|x-x'|^{(m+\frac 1m)/2}},
\eqno (4.8)
$$
which is the contribution to the   Luttinger fermion one-particle
density matrix from one of the two fermi points.

Given these two  results, it seems clear that our method of
treating the integrals over the electron coordinates must be
equivalent to the boson field theory manipulations of section 3.
The connection is made by reviewing the derivation of the
Luttinger model wavefunction presented in the appendix. To obtain
the wavefunction we integrate over all the boson modes except for
those on a single time slice. The functional integration over
$\chi$ in (4.5)  is the boson field integration over this final
time slice.

\vskip 40pt
\line {\bf 5. Discussion \hfil }

The principal result of this paper is a prescription for
extracting the Luttinger liquid picture of the FQHE edge states
directly  from the Laughlin wavefunction: We first dimensionally
reduce the wavefunction by constraining  all its arguments to lie
on the boundary of the two-dimensional electron gas. Then,
motivated by the fact that the electrons are  really some
distance within the system, we smooth out the charges in the
resulting one-dimensional Coulomb plasma so that the operators can
distinguish between the passage of a particle that gives an
$e^{i\pi}$ phase from the passage of one that gives an
$e^{3i\pi}$ phase. This  crucial step is the only relic of the
direction perpendicular to the edge.  Once we have performed
these operations, we can use the Luttinger liquid {\it quantum
placet.}

A comment about the resulting statistical  transmutation is in
order. Part of the lore of the FQHE is that electrons in Laughlin
states bind to an even number of vortices which serve to
transform them into hyper-fermions [20]. This is a dynamical
effect.  The operators $\psi$, $\psid$ we use to create and
annihilate the electrons are simple unadorned Fermi fields with
conventional commutation relations. Since the electrons abandon
their vortices as soon as they leave the FQHE system, it is the
density matrix and other Green functions  defined in terms of
these conventional statistics operators that govern the
interaction of the system with the outside world, and it is these
we compute.   When we examine the  density matrix in the bulk we
see no sign (beyond a reduced density) of the statistical
dressing. Only as we approach the boundary of the electron gas
does it begin to display the effects of the attached vortices.
At the boundary, the density matrix, and indeed all
time-dependent, conventional-statistics Green functions coincide
with  one-dimensional Luttinger liquid correlators of enhanced
statistics operators.

{}From the Coulomb gas analysis of the last section we begin to
understand why we need the boundary to see the effects of the
dynamical  statistics change. Only near a boundary do we tend to
have more electrons on one side than the other, and thus an
opportunity  to perceive the accumulating phase change.

The insight that comes  from knowing how to implement the
connection between the bulk wavefunction and the Luttinger liquid
behaviour in the simple Laughlin states should be useful for
understanding the connection between the edge  behaviour of more
complicated fractional phases and their candidate wavefunctions.

\vskip 40pt
\line{\bf Acknowledgements\hfil}

This work was supported by the National Science Foundation under
grant numbers  PHY89-04035 and DMR91-22385.  We would like to
thank Jainendra Jain and Rajiv Singh for showing us ref. [17],
and asking the questions that lead to this work. We  also thank
Charlie Kane for illuminating conversations.

\vskip 40pt
\line{\bf Appendix\hfill}

In this appendix we will derive the many-body Luttinger-Thirring
ground-state wavefunction
$\Psi(\theta_1,\theta_2,\ldots,\theta_N)$ for $N$ fermions living
on a circle of circumference $2\pi$. We will use a slightly
modified form of the methods in ref.  [12]. It is most convenient
to use the Lorentz invariant Thirring form  for this.  We begin
by finding  the ground-state wavefunction $\Phi$ for the
bosonized version.  Because the  bosonization of the Thirring
model is most familar from the work of Coleman [19] we will use
his conventions in this appendix. The boson fields here therefore
differ in normalization by a factor of $2\sqrt {\pi}$ from those
in the main text.

The bosonized action is
$$
S=\frac {4\pi}{\beta^2}\int \frac 12 (\partial \varphi)^2 d\tau
d\theta
\eqno (A.1)
$$
Here $\beta^2$  is Coleman's parameterization of the interaction.
For the free theory $\beta^2=4\pi$. To relate it to other
paramterizations of the interaction it is best to compare
correlators calculated in the different schemes. We find
$4\pi/\beta^2=\lambda=e^{-2\alpha}$.

In a Schr{\"o}dinger representation  the wavefunction
$\Phi(\varphi_c)$ is a functional of the boson field
configuration. To compute it we  take a path integral  over
$\varphi$'s defined on the half-cylinder $\Omega=[-\infty,
0]\times S^1$, where the argument of the wavefunction,
$\varphi_c$, appears as the boundary condition
$\varphi_c(\theta)=\varphi(0,\theta)$ imposed on the circle at
$\tau=0$. The long euclidean time interval between $-\infty$ and $0$
projects out the ground state, and  the unnormalized wavefunction
is  given by
$$
\Phi(\varphi_c)= \langle vac \vert \varphi_c\rangle  =\int_0^{\varphi_c}
d[\varphi] e^{ -\frac {4\pi}{\beta^2}\int_{\Omega} \frac 12 (\partial
\varphi)^2 d\tau d\theta}.
\eqno (A.2)
$$
Being quadratic, the path integral may be performed by replacing the
$\varphi$ in the integrand with the solution to Laplace's
equation for the given $\varphi_c$ boundary values. Using the
standard formula for the solution to the Dirichlet problem,
$$
\varphi(\tau',\theta')=\oint d\theta
\varphi_c(\theta)\partial_\tau
G_{\Omega} (\tau,\theta;\tau'\theta')\vert_{\tau=0},
\eqno (A.3)
$$
and integrating by parts, we find that the exponent
$$
E=\int_{\Omega} d\tau d\theta \frac 12 (\partial\varphi)^2
\eqno (A.4)
$$
can be written in terms of the boundary data,
$$
E=\frac 12 \oint d\theta \varphi_c(\theta)\partial_\tau
\varphi(0,\theta)
=
-\frac12
\int_{S^1}\int_{S^1}d\theta d\theta'
\varphi_c(\theta)\varphi_c(\theta')\partial_\tau\partial_{\tau'}
G_{\Omega}(\tau,\theta,\tau',\theta').
\eqno (A.5)
$$

In these formulae $G_{\Omega}$ is the Dirichlet Green-function on
the half cylinder, {\it i.e.}
$$
\nabla^2_{\bf r}G_{\Omega}(\r,\r')=\delta^2(\r-\r')
\eqno (A.6)
$$ and
$G_{\Omega}(\r,\r')=0$ if  $\r$ is on the boundary circle.

The   Green function $G$ for the infinite cylinder is obtained
as
$$
G(\r,\r')=-\frac 1{2\pi}{\rm Re\,}\ln(e^{iz}-e^{iz'})= -\frac
1{2\pi}\ln\vert \sin(z-z')/2\vert,
\eqno (A.7)
$$
with $z=\tau+i\theta$, by a
conformal transformation of  the ${\bf R}^2$ Green function,
$G_0(\r,\r')=-\frac 1{2\pi}\ln|\r-\r'|$. The half cylinder Green function,
$G_{\Omega}$, is then found from $G$ by the method of images:
$$
G_{\Omega}(\tau,\theta,\tau',\theta')=
G(\tau-\tau',\theta-\theta')-G(\tau+\tau',\theta-\theta').
\eqno (A.8)
$$

We now use  Laplace's equation and the form of the arguments of
$G$ to trade the partial derivatives with respect to $\tau$ for
partials with respect to $\theta$
$$
\partial_\tau\partial_{\tau'}G_{\Omega}(\tau,\theta,\tau'\theta')=
-\partial_\theta\partial_{\theta'}G(\tau-\tau',\theta-\theta')
-\partial_\theta\partial_{\theta'}G(\tau+\tau',\theta-\theta').
\eqno (A.9)
$$
We need this expression only on $\tau=\tau'=0$ where it is equal to
$-2\partial_\theta\partial_{\theta'}G(0,\theta-\theta')$.

After a final integration by parts, the fruit of our labours is
$$
E=\frac 12 \int\int d\theta d\theta'
\partial_\theta\varphi_c(\theta)\partial_{\theta'}\varphi_c(\theta')
\frac 1{\pi} \ln\vert \sin(\theta-\theta')/2\vert,
\eqno (A.10)
$$
and the exponential of this $\Phi=\exp (-4\pi E/\beta^2) $ is the harmonic
oscillator like ground state of the Bose field.

We convert $\Phi$  to a many-body Fermi wavefunction $\Psi$ expressed in terms
of the
particle locations by using the bosonization rule $\rho=\frac
1{\sqrt{\pi}}\partial_\theta \varphi$, and replacing the density
with its first quantized form $\rho(\theta)=\sum_{i=1}^N
\delta (\theta-\theta_i)$. We find that
$$
\Psi(\theta_1,\theta_2,\ldots,\theta_N)=
|e^{-i(N-1)\sum \theta_i}\prod_{i<j}( e^ { i
\theta_i}-e^{i \theta_j})|^{4\pi/\beta^2}
\eqno (A.11)
$$
Since we only know that there is  some particle at the location
$\theta_i$, and not which particular particle it is, this expression
is to be used for the standard  ordering $\theta_1<\theta_2\ldots
<\theta_N$ only. The values of $\Psi$ for other orderings of the
arguments are  found by imposing  the antisymmetry.

\vskip 40pt
\line{\bf References\hfil}

\item{[1]} X.~G.~Wen, Phys. Rev. Lett. {\bf 64} (1990) 2206;
Phys. Rev.  {\bf B41} (1990) 12838; {\bf B43} (1991) 11025; {\bf
B44} (1991) 5708.

\item{[2]} K.~Moon, H.~Yi, C.~L.~Kane, S.~M.~Girvin,
M.~P.~A.~Fisher, Phys. Rev. Lett. {\bf 71} (1993) 4381.

\item{[3]}  C.~L.~Kane, M.~P.~A.~Fisher, Phys. Rev. Lett.
{\bf 68} (1992) 1220; Phys. Rev. {\bf B46} (1992) 7268, 15233.

\item{[4]} F.~Milliken, C.~P.~Umbach,  R.~A.~Webb, IBM preprint.

\item{[5]} F.~D.~M.~Haldane, Phys. Rev. Lett. {\bf 45} (1980)
1358; J. Phys. {\bf C14} (1981) 2585.

\item{[6]} S.~Mitra, A.~H.~Macdonald, Phys. Rev. {\bf B48} (1993)
2005.

\item{[7]} X.~G.~Wen, Int. J. Mod. Phys. {\bf B6} (1992) 1711.

\item{[8]} R.~B.~Laughlin, Phys. Rev. Lett {\bf 50} (1983)
1395.

\item {[9]} S-C.~Zhang, Int. J. Mod. Phys {\bf 6} (1992) 25.

\item {[10]} A.~Karlhede, S.~A.~Kivelson, S.~ L.~ Sondhi, in
{\it Correlated Electron Sytems}\/  V.~J.~Emery ed. (World
Scientific 1993).

\item{[11]} A.~Lopez, E.~Fradkin, Phys. Rev. Lett. {\bf 69} (1992)
2126.

\item{[12]} E.~Fradkin, E.~Moreno,
F.~A.~Schaposnik, Nucl. Phys. {\bf B392} (1993) 667.

\item{[13]} B.~Sutherland, J. Math. Phys. {\bf 12} (1971) 246;
Phys. Rev {\bf A 4} (1971) 2019; {\bf A5} (1972) 1372.

\item{[14]} N.~Kawakami, S-K.~Yang, Phys. Rev. Lett. {\bf 67}
(1991) 2493.

\item{[15]}M.~Stone, Ann. Phys. {\bf 207} (1991) 38;
                    Phys. Rev. {\bf B42} (1990) 8399;
                    Int. J. Mod. Phys. {\bf 5} (1991) 509.

\item{[16]} F.~D.~M.~Haldane, Phys. Rev. Lett. {\bf 47} (1981)
1840.

\item{[17]} C.~S.~Hellberg, E.~J.~Mele, Phys. Rev. Lett. {\bf 68}
(1992)  3111.

\item{[18]} J.~Martinez, M.~Stone, Int. J. Mod. Phys. {\bf B7}
(1993) 4389.

\item{[19]} S.~Coleman,  Phys. Rev. {\bf D11} (1975) 2088.

\item{[20]}  B.~I.~Halperin, Helvetica Phys. Acta, {\bf 56}
(1983) 75.

\bye